\documentclass[prc,aps,twocolumn,floatfix]{revtex4}

\usepackage{graphicx}

\begin{document}

\title{Hadronic Signals of Deconfinement at RHIC}
\author{Berndt M\"uller}
\affiliation{Department of Physics, Duke University,
             Durham, NC 27708-0305, USA}

\begin{abstract}
This article reviews (soft) hadronic signals of deconfinement
and chiral symmetry restoration in hot QCD matter in the light
of the results from the first three years of the experimental
program at the Relativistic Heavy Ion Collider. 
\end{abstract}
\maketitle

\section{Introduction}

Quantum mechanics dictates that even ``empty'' space is not 
empty, but rather filled with quantum fluctuations of all 
possible kinds. In many contexts, such as in atomic physics, 
these vacuum fluctuations are subtle effects which can only 
be observed by precision experiments. In other situations, 
especially when interactions of sufficient strength are involved, 
the vacuum fluctuations can be of substantial magnitude and 
even ``condense'' into a nonvanishing expectation value of some
quantum field. These vacuum condensates can act as a medium
\cite{Le80}, which influences the properties of particles 
propagating through it. 

An important example of such a vacuum condensate is the Higgs 
vacuum, which is introduced in the Standard Model of particle 
physics to generate the masses of quarks, leptons, and the gauge 
bosons of the weak interaction. The vacuum expectation value of 
the Higgs field, $\langle\phi\rangle = 246$ GeV, is uniquely 
determined in the Standard Model; the quark and lepton masses 
differ from one another only due to the different strength of 
the coupling of each fermion to the Higgs field. The quark masses 
receive additional contributions from the quark and gluon 
condensates in the QCD vacuum, which have been measured to be:
\begin{eqnarray}
\langle{\bar q}q\rangle = (235 {\rm MeV})^3 , \nonumber \\ 
\langle g^2G_{\mu\nu}G^{\mu\nu}\rangle = (840 {\rm MeV})^4 .
\label{eq01}
\end{eqnarray}
In fact, the contribution of the QCD vacuum condensates to the
masses for the three light quark flavours $u, d, s$ considerably
exceed the mass believed to be generated by the Higgs field
{see Fig.~\ref{fig1}).

\begin{figure}   
\includegraphics[width=\linewidth]{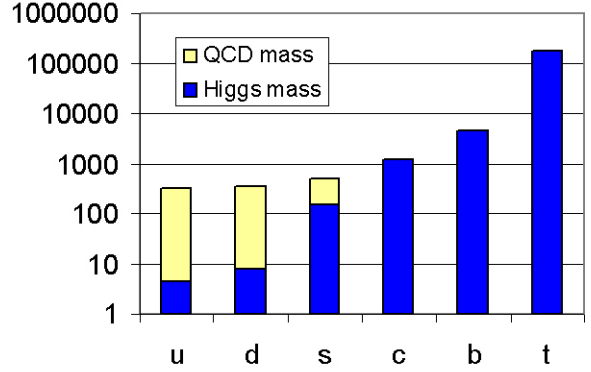}
\caption{Masses of the six quark flavors. The masses generated by
electroweak symmetry breaking (current quark masses) are shown in
dark blue; the additional masses of the light quark flavors generated 
by spontaneous chiral symmetry breaking in QCD (constituent quark 
masses) are shown in light yellow. Note the logarithmic mass scale.}
\label{fig1}
\end{figure}

If the vacuum acts as a medium and influences the properties
of fundamental particles and their interactions, its properties
can conceivably change. This idea has important implications in
cosmology (inflation \cite{inflation}) and in modern versions
of the anthropic cosmological principle \cite{Re03,Da04}, which
invoke the concept of multiple universes with different 
phenomenological properties due to the realization of different 
vacuum states. The most promising candidate for the fundamental 
theory of space-time, superstring theory, promises to sustain 
an almost unlimited number of vacuum states \cite{Su03}, lending 
some credence to this concept. These theoretical speculations 
call for an experimental verification of the existence of multiple 
vacuum states and the possibility of transitions among them under
the influence of external constraints.

Because of the energy scales involved, only the QCD vacuum is 
amenable to modification at energies accessible with present 
technologies. The Relativistic Heavy Ion Collider (RHIC) was 
built to explore these questions by creating conditions in
the laboratory under which the structure of the QCD vacuum 
would change and the properties of particles and forces of the 
strong interactions would be modified. The beam energy of RHIC 
was chosen to be high enough to generate temperatures commensurate 
with the energy scale of the QCD vacuum condensates and thus to 
affect the structure of the QCD vacuum. 

Numerical simulations of lattice QCD predict that a dramatic 
change in the QCD vacuum state occurs around the critical 
temperature $T_c \approx 160$ MeV, where the quark and gluon 
condensates melt, and the vacuum takes on a simpler structure. 
The part of the light quark masses that is induced by the 
quark condensate disappears above $T_c$ and only the much 
smaller masses generated by the electroweak Higgs field remain. 
The degrees of freedom corresponding to independently propagating 
gluons, which are frozen in the normal QCD vacuum, are also 
liberated above $T_c$. This pattern is visible in Fig.~\ref{fig2}, 
which shows the dramatic jump in the scaled energy density 
$\epsilon(T)/T^4$ at the critical temperature. 

\begin{figure}   
\includegraphics[width=\linewidth]{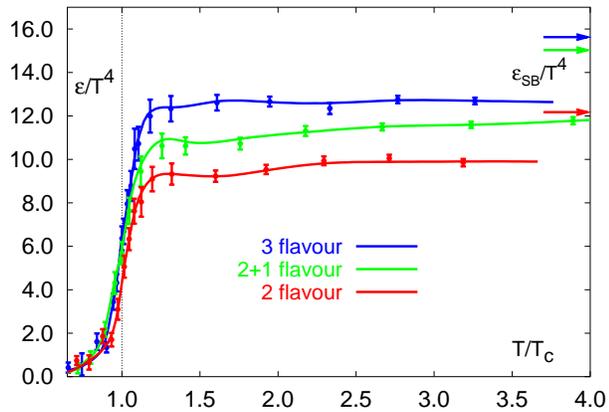}
\caption{Scaled energy density $\epsilon/T^4$ for thermal lattice-QCD
with two and three light quark flavors and for two light and one heavier
flavor (from Karsch \protect\cite{Ka00}).}
\label{fig2}
\end{figure}

\section{Hadronic probes of deconfinement}

Which observables allow us to confirm that the predicted 
transformation of the QCD vacuum actually occurs? Here we
focus on signals involving light ($u, d, s$) quarks, the 
primary constituents of the medium. This restriction is not 
intended to minimize the importance of other observables,
such as hard probes, covered in different contributions 
to this volume \cite{others}. In fact, the strong quenching
of the emission of hadrons at large transverse momenta 
and the existence of a strong (anisotropic) collective flow
are important prerequisites for the apparent dominance of 
quark recombination at momenta of a few GeV/c, which will
be discussed below.

Two effects caused 
by the disappearance of the vacuum condensates stand out: 
The melting of the quark condensate lowers the effective 
mass of the $s$-quark from about 500 MeV to less than 150 MeV, 
making it easy to create $s$-quark pairs in copious quantities 
\cite{Ra82,RM82,KMR86}. The dissolution of the gluon condensate 
and the concomitant screening of the long-range color force
by thermal gluons allows light quarks to propagate outside of 
hadronic bound states. As the hot quark-gluon plasma (QGP) cools,
hadrons should form by recombinant emission of quark-antiquark
pairs (mesons) or color-singlet triplets of quarks (baryons)
from the surface of the hot medium \cite{Vi91}, preserving
the collective flow pattern of their partonic constituents
\cite{Vo03}.

\subsection{Flavor equilibration}

Let us consider these arguments more quantitatively. The
expectation that the formation of a deconfined quark-gluon 
plasma results in a substantial enhancement of the production
of hadrons carrying strangeness is based on a comparison of 
energy thresholds and rates. In a hadronic environment, the
energy required to excite strangeness is about 700 MeV, as
in the reaction $\pi\pi\to K{\bar K}$, or about $4T_c$. 
In the QGP, the threshold for the dominant reaction 
$gg\to s{\bar s}$ is about 300 MeV, less than $2T_c$.
Hadronic reactions producing multiply strange hadrons, such
as the $\Xi$ and $\Omega$ hyperons, are thus predicted to 
be slow under thermal equilibrium conditions and not leading
to hadrochemical equilibrium within the time available in a
heavy-ion reaction \cite{KR84}. On the other hand, flavor
equilibrium among $u, d, s$ quarks is predicted to develop
within a QGP on the available time scales \cite{RM82}.
Hadrons containing any number of strange quarks should, 
therefore, be produced in equilibrium abundances when a QGP
is formed.

The prediction of a strong enhancement of the production
of multi-strange baryons and antibaryons is independent
of the details of the hadronization model, although the
quantitative predictions for individual hadron yields may
differ slightly \cite{BZ83,KMR86,Ba88,Bi95,RL00}. Extensive
studies of final state effects on the chemical composition
of the emitted hadrons have concluded that the abundances
of multi-strange baryons are only slightly affected by
inelastic rescattering \cite{BD00}. The consistency of this 
scenario was confirmed recently once more in a systematic 
study of baryon pair production based on model independent 
rate equations, which showed that a strong enhancement of 
multi-strange baryon production requires the assumption of 
flavor equilibrium already at the start of the hadronic 
phase evolution \cite{KS03,HK04}.

It is important to distinguish the chemical equilibrium
distribution attained in a deconfined plasma phase from the 
statistical distribution of hadrons produced in $e^+e^-$ or 
p+p reactions\cite{Hage,Be96,BH97}. In those reactions, the
production of hadrons carrying strange quarks is systematically
suppressed. Although this phenomenon can be explained as an 
effect of flavor conservation in a small volume of hadronic 
size, it is precisely the {\em disappearance} of this suppression 
in heavy-ion collisions, corresponding to the transition from 
a canonical ensemble to a macrocanonical one, which is the 
signature of the creation of locally deconfined QCD matter.

The observation of a strong enhancement of multi-strange
baryon and antibaryon emission in Pb+Pb collisions by
experiment WA97 \cite{WA97} formed an important pillar of 
the claim that evidence for a new state of matter had been 
found at the CERN-SPS \cite{HJ00}. The results were 
confirmed and extended to lower beam energies in recent 
years by the NA57 collaboration \cite{NA57}. The conclusion 
that a deconfined phase is formed under the conditions 
created by Pb+Pb collisions at CERN-SPS energies is also 
supported by a recent comprehensive analysis of the data 
for several relevant observables in the framework of a 
schematic, but realistic model of the space-time
evolution of the collisions \cite{Re04}.

As impressive as these results are, the evidence remains
indirect and relies on the argument that a hadrochemical
equilibrium cannot be established by interactions among
hadrons with confined quarks. The higher beam energies
available at RHIC have made it possible to explore regions
of phase space where the formation of hadrons from
recombining quarks can be directly observed and the partonic
origin of their collective flow pattern can be established.
It is important to recognize that, in contrast to the
strangeness enhancement itself, these phenomena were not
quantitatively predicted before the observations were made.
However, I will argue that the theoretical arguments, even
if they were developed under the motivation of the first 
experimental results from RHIC, are sufficiently robust to 
provide compelling evidence for the formation of a deconfined 
quark phase, as more complete and statistically precise data 
are becoming available.

\subsection{Quark recombination}

Let us begin with a discussion of quark recombination
(see Fig.~\ref{fig3}a).
The statistical coalescence of two or more particles to form
a bound state is a widely used concept, which remains somewhat 
murky in many cases for two reasons. First, the formation of 
a bound state requires that either at least one of the two 
particles is off-shell or that another particle carries away 
the surplus energy. Second, the implicit reduction in the 
number of degrees of freedom during coalescence
tends to conflict with the second law of thermodynamics, which 
dictates that the entropy cannot decrease. Fortunately, there
are regions of phase space -- hadrons emitted with highly
epithermal momenta -- where these concerns do not constitute 
serious obstacles. Hadrons emitted with a high momentum make 
an almost instantaneous transition from the dense medium into 
the surrounding vacuum. Another way to state this is to say that 
-- observed from the fast moving hadron -- the radiating
surface appears highly Lorentz contracted and thus thin on a
hadronic length scale. Energy conservation is therefore not an 
important constraint on the recombination process. Moreover, at
high transverse momentum, the masses of quarks and hadrons can
be neglected in first approximation with corrections of order
${\cal O}(m^2/p_T^2) \ll 1$. And finally, fast hadrons constitute
only a very small fraction of all emitted particles, thus 
limiting possible violations of the second law of thermodynamics.

\begin{figure}   
\begin{center}
\includegraphics[width=0.4647\linewidth]{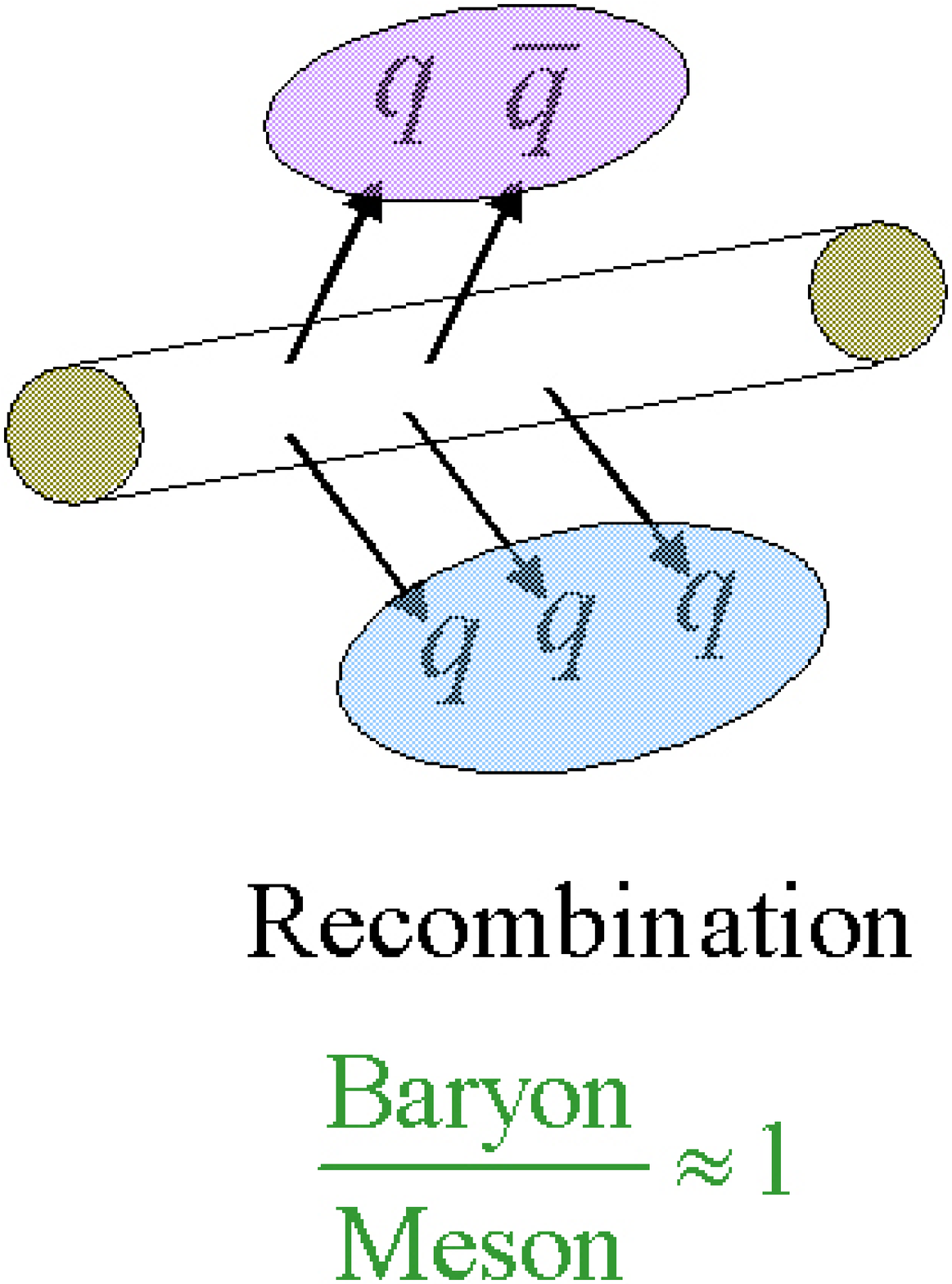}
\hspace{0.05\linewidth}
\includegraphics[width=0.4353\linewidth]{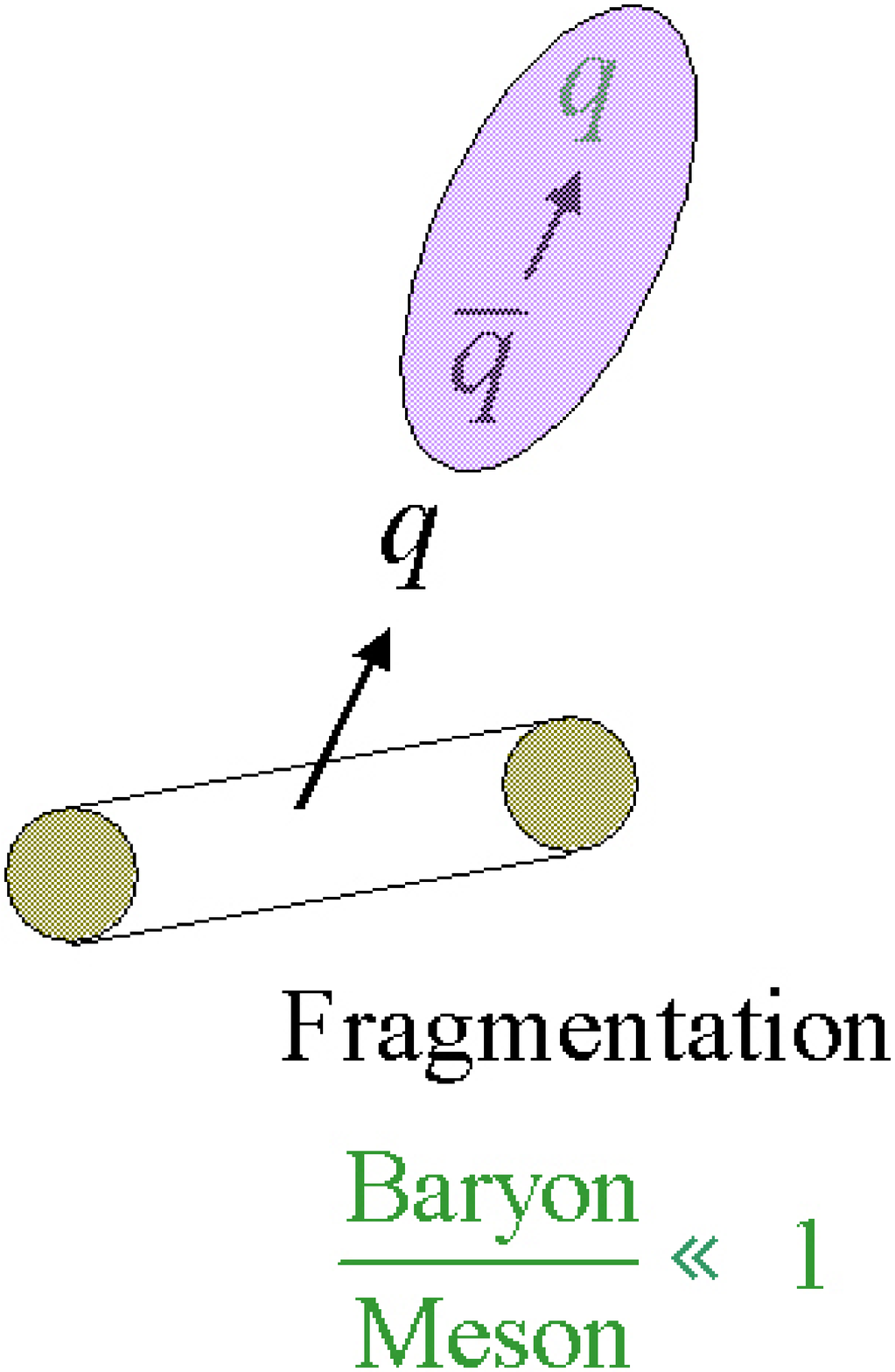}
\end{center}
\caption{(a) Left panel: Mesons and baryons formation by
recombination of quarks and/or antiquarks from a thermal,
deconfined medium predicts a baryon/meson ratio of order
unity. (b) Right panel: Hadron production by fragmentation
of an energetic parton favors meson over baryon production.}
\label{fig3}
\end{figure}

Denoting the quark phase-space distribution by $w_{\alpha}(p)$ 
(here $\alpha$ denotes color, flavor, and spin degrees of freedom) 
and employing light-cone coordinates along the direction of 
the emitted hadron, instantaneous recombination predicts the 
following meson and baryon spectra, respectively \cite{Fr03}:
\begin{widetext}
\begin{equation}
E\frac{dN_M}{d^3P} = \int d\Sigma \frac{P\cdot u}{(2\pi)^3} 
\sum_{\alpha\beta} \int dx\, 
w_{\alpha}(xP^+) {\bar w}_{\beta}((1-x)P^+) 
|\phi^{(M)}_{\alpha\beta}(x)|^2, 
\label{eq02}
\end{equation}
\begin{equation}
E\frac{dN_B}{d^3P} = \int d\Sigma \frac{P\cdot u}{(2\pi)^3} 
\sum_{\alpha\beta\gamma} \int dx\, dx'\, 
w_{\alpha}(xP^+) w_{\beta}(x'P^+) w_{\gamma}((1-x-x')P^+) 
|\phi^{(B)}_{\alpha\beta\gamma}(x,x')|^2.
\label{eq03}
\end{equation}
Here $\Sigma$ stands for the freeze-out hypersurface, 
$u^\mu$ for the collective flow velocity of the partonic 
matter, and capital letters $P$ denote hadron coordinates. 
For a thermal parton distribution, $w(p) = \exp(p\cdot u/T)$, 
the partonic factors in both equations combine to a single 
thermal distribution for the hadron, $w(P^+)$. In this case 
the integrations over the momentum fractions $x, x'$ become
trivial, and the end result is independent of the details of 
the hadron wavefunction $\phi$. The prediction is that hadrons 
are emitted in thermal and chemical equilibrium abundances and 
acquire the collective radial flow velocity of the quarks. In
fact, one can argue that the recombination mechanism provides
the justification for applying the statistical model to
describe ratios of hadron yields at up to momenta of several
GeV/c.

In the absence of a thermal medium, energetic hadrons are 
created by the fragmentation of a fast quark or gluons
(see Fig.~\ref{fig3}b).
This mechanism predicts a hadron spectrum of the form
\begin{equation}
E\frac{dN_h}{d^3P} = \int d\Sigma \frac{P\cdot u}{(2\pi)^3} 
\sum_{\alpha} \int dz\,z^{-3}\, w_{\alpha}(P^+/z) 
D_{\alpha\to h}(z) ,
\label{eq04}
\end{equation}
\end{widetext}
where $z<1$ denotes the momentum fraction carried by the hadron. 
It is easily shown that for large values of the momentum $P^+$ 
recombination always wins over fragmentation in the case of an
exponentially falling momentum distribution, because $w(P^+/z)$ 
falls off faster than $w(P^+)$. On the other hand, fragmentation 
asymptotically dominates over recombination, when the parton 
distribution obeys a power law as it is always the case in QCD 
at sufficiently large momentum. In relativistic heavy ion 
collisions the power law tail is suppressed (in central Au+Au 
collisions at RHIC by about a factor 5) due to the energy loss 
of fast partons during their passage through the hot, dense 
matter. This mechanism delays the dominance of fragmentation 
over recombination to rather large transverse momenta, about 
4 GeV/c for mesons and 6 GeV/c for baryons. This is illustrated
in Fig.~\ref{fig4}, which shows the two contributions to the
pion spectrum in central Au+Au collisions at RHIC separately,
together with the PHENIX data.

\begin{figure}   
\begin{center}
\includegraphics[width=\linewidth]{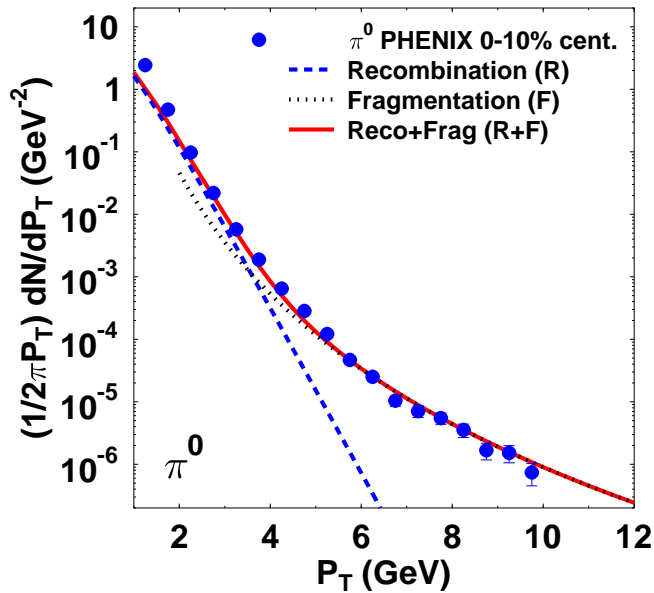}
\end{center}
\caption{Spectrum of pions emitted in central Au+Au collisions
at RHIC for $\sqrt{s_{\rm NN}}=200$ GeV. The contributions from
recombination and fragmentation are shown separately, together
with the data from the PHENIX collaboration 
(from \protect\cite{Fr03}).}
\label{fig4}
\end{figure}

\subsection{Collective flow of quarks}

When the radial flow velocity of the QGP exhibits an azimuthal 
anisotropy, as in noncentral collisions of two nuclei, this
anisotropy becomes imprinted on the hadrons formed by quark
recombination. If the quark flow anisotropy has the form 
$[1+v_2(p)\cos 2\varphi_p]$, where $\varphi_p$ is the emission 
angle with respect to the collision plane and $v_2(p)$ the 
so-called elliptic flow velocity, the elliptic flow parameter 
for a meson or baryon is (in the usual case where $v_2\ll 1$):
\begin{eqnarray}
v_2^{(M)}(P) &=& v_2(xP) + v_2((1-x)P) , \\
v_2^{(B)}(P) &=& v_2(xP) + v_2(x'P) + v_2((1-x-x')P) \nonumber .
\label{eq05}
\end{eqnarray}
In the limiting case of equal momentum fractions carried by
the constituent quarks these relations simplify to 
\cite{Vo03,MV03,Fr03}
\begin{eqnarray}
v_2^{(M)}(P) &\approx & 2 v_2(P/2) , \nonumber \\
v_2^{(B)}(P) &\approx & 3 v_2(P/3) .
\label{eq06}
\end{eqnarray}
In the hydrodynamic regime, where the flow anisotropy grows
roughly linearly with the momentum, $v_2(p) = \alpha p$, the 
scaling factors cancel, and mesons and baryons exhibit an equal 
$v_2$ at the same momentum. At large transverse momentum, however,
where the anisotropy is limited by transparency effects, this
is no longer the case, and a collective anisotropic flow carried 
by quarks can be distinguished from one carried by a hadronic 
fluid. Final state effects on the elliptic flow in the hadronic
phase have been shown to be small \cite{LK02}.

\subsection{Unified descriptions}

These considerations based on general principles can be 
extended in several ways to cover a wider range of hadron 
momenta \cite{Fr04}. By choosing the medium rest frame instead
of the light-cone frame of the emitted hadron, one can obtain
a more traditional coalescence model description of the hadron 
spectrum down to $p_T=0$ \cite{Gr03}. The price one pays for 
this extended coverage is the need to make specific assumptions 
about the internal wavefunction of the emitted hadrons, which 
are by necessity simplistic. A virtue of this approach
is that one can incorporate final-state decays of unstable
hadrons, such as the $\rho$-meson, which help to resolve the
entropy problem and improve the description of pion spectra,
especially at small momenta \cite{Gr04}. Effects due to hadron 
masses can also be treated more consistently in such models.

Another extension concerns the treatment of fragmentation
processes. In the presence of a medium, a fast parton may form
a hadron by picking up one or more constituents from the medium.
This process has been called fast-slow recombination \cite{Gr03}.
Depending on the hadron wavefunction, it may contribute to
hadron emission up to quite large transverse momenta extending
into the $6-8$ GeV/c range. Alternatively, one can consider the
process of hadron formation via fragmentation also as a 
recombination process acting on a primary distribution of
fragmented shower partons \cite{DH77,HY02}. This approach 
gives a microscopic description of the fragmentation functions 
$D_{\alpha\to h}(z)$, allowing one to extend the calculation 
consistently, albeit model dependently, to hadron production 
by recombination involving partons from a thermal medium 
\cite{HY04}. This model predicts a large baryon-to-meson 
ratio in the $p_T$-range of $3-8$ GeV/c, where the mixed 
recombination of thermal and shower partons dominates. This
approach is also aplicable to hadron production in p+p and
d+Au collisions, where it provides an explanation for the
differences in the Cronin effect observed for mesons and 
baryons \cite{HY04b}.

\subsection{Principal signatures}

These considerations suggest several critical signatures of 
the formation of a thermal quark-gluon plasma accompanied by
a structural change in the QCD vacuum: 
\begin{itemize}
\item
Hadrons containing any number of strange quarks should be 
produced in near equilibrium abundances; 
\item
Hadrons created by recombination of moderately energetic 
quarks should form the dominant mode of hadron emission at
low and intermediate momenta;
\item
Baryons should be produced as abundantly as mesons at
intermediate transverse momenta;
\item
The collective flow patterns of the emitted hadrons should
reflect the collective flow of their constituent quarks.
\end{itemize}
As we will discuss in the next section, these characteristic 
phenomena have been observed in Au+Au collisions at RHIC.

\section{Results from RHIC}

Fig.~\ref{fig5} shows that hadrons containing any number
of $s$-quarks are produced according to the expectation 
that a chemically equilibrated quark-gluon plasma converts 
into hadrons maintaining flavor equilibrium near the critical 
temperature. The value deduced from the data ($T_{\rm ch} = 
177$ MeV \cite{PBM}) is equal to $T_c$ within the present 
theoretical uncertainties. There exist significant deviations
from the predicted equilibrium abundances for some short-lived
hadronic resonances, such as the $\Delta(1232)$, $K^*$, and
the $\Lambda(1520)$. These deviations can be qualitatively
understood as final state effects due to medium modifications
of the resonance spectral function or absorption and
regeneration effects \cite{Zh04}.

\begin{figure}   
\includegraphics[width=\linewidth]{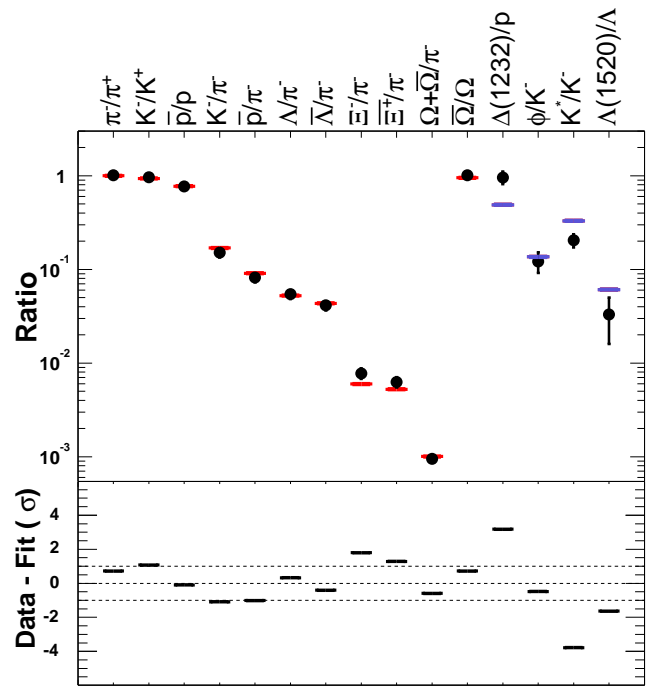}
\caption{Comparison of the ratios of observed yields of various
hadrons at midrapidity in Au+Au collisions at RHIC with the
ratios predicted in a thermal and chemical equilibrium model
(from STAR).}
\label{fig5}
\end{figure}

The apparent lack of a nuclear suppression of protons 
\cite{PHENIX-p} and $\Lambda$-hyperons \cite{STAR-L} 
in Au+Au collisions in the transverse 
momentum range $2-4$ GeV/c, where pions are already 
suppressed by a factor $4-5$, came as a surprise. The 
measurements have been extended to other identified mesons
and baryons, including the $\phi$ and the $\Xi$ (see Fig.\
\ref{fig6}), confirming the baryon-meson dichotomy which 
is a generic feature of the recombination mechanism. The 
observed onset of nuclear suppression for the baryons between 
4 and 6 GeV/c is in excellent agreement with the predictions 
by the Duke and Texas A{\&}M models (see Fig.~\ref{fig7})
and reflects the transition from a recombination dominated
to a fragmentation dominated hadronization regime. The
fall-off of the baryon/meson ratio at small $p_T$ (see
Fig.~\ref{fig7}a) is due to mass effects.

\begin{figure}
\begin{center}
\includegraphics[width=0.98\linewidth]{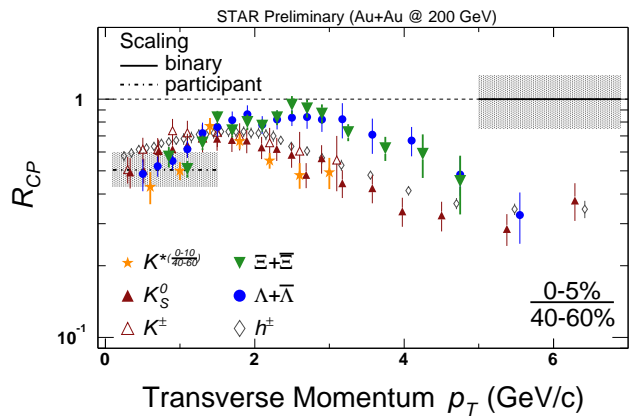}
\includegraphics[width=\linewidth]{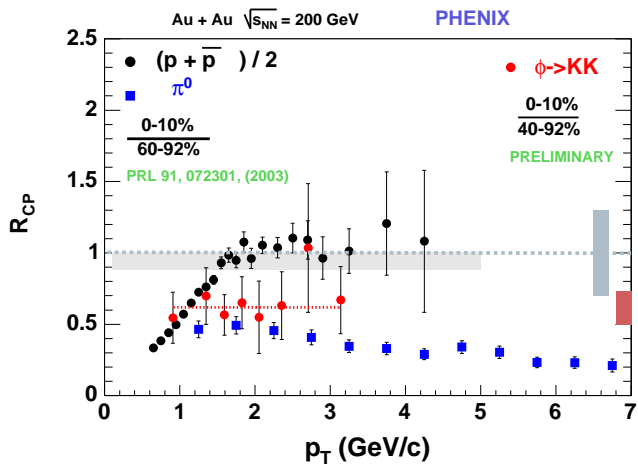}
\end{center}
\caption{Nuclear suppression observed in central (compared
with peripheral) Au+Au collisions for various hadrons. 
(a) Upper panel: STAR \cite{STAR-QM04}; 
(b) lower panel: PHENIX \cite{PHENIX-QM04}.}
\label{fig6}
\end{figure}

Measurements of the flow patterns of the emitted hadrons 
provide essential additional evidence for the hypothesis that 
they are created directly from a phase of unconfined quarks and 
antiquarks. The experiment makes use of the fact that the
region of heated vacuum in semiperipheral collisions between 
two nuclei is almond-shaped and thus sustains anisotropic
pressure gradients. In the hydrodynamic limit this leads to 
an anisotropic expansion pattern called ``elliptic flow''
\cite{Ol92}, which is characterized by the parameter $v_2$. 
As already mentioned, the flow anisotropy is an increasing 
function of the momentum of the emitted particle in the
hydrodynamic regime and saturates when transparency effects
set in at large transverse momenta. 

\begin{figure}
\begin{center}
\includegraphics[height=0.9\linewidth,angle=-90]{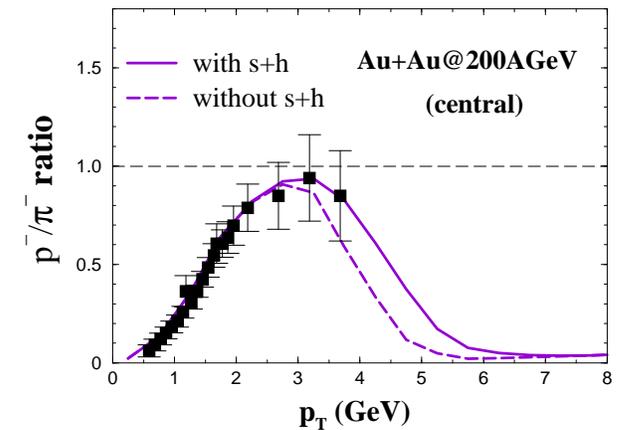}
\includegraphics[width=\linewidth]{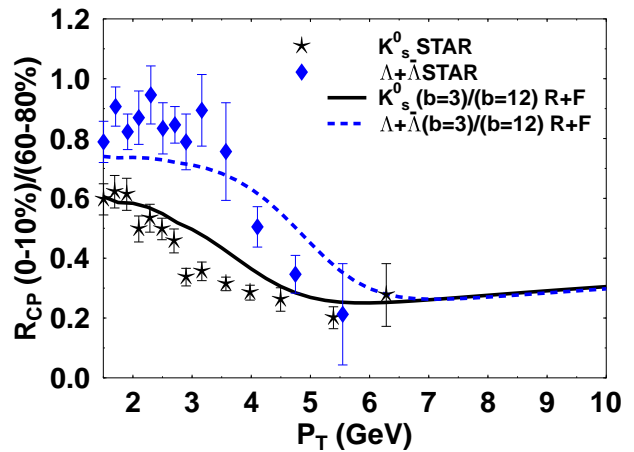}
\end{center}
\caption{Baryon enhancement as function of transverse momentum
predicted by two different recombination models, in comparison
with RHIC data. (a) Upper panel: Texas A{\&}M model for $p/\pi$.
(b) Lower panel: Nuclear suppression of $\Lambda$-hyperons and 
kaons in the Duke model.}
\label{fig7}
\end{figure}

The RHIC experiments 
have confirmed this expected behavior, but found that it 
varies sustantially from one hadron species to another 
(see Fig.~\ref{fig8}a). However, when the constituent
rescaling (\ref{eq06}) is applied to these data, the flow 
patterns are found to collapse onto a common line (see 
Fig.~\ref{fig8}b), suggesting that one observes the flow 
pattern of individual quarks, which coalesce into hadrons 
only at the end of the expansion \cite{STARv2}. The onset
of saturation at a quark momentum $p_T\approx 1$ GeV/c is 
consistent with the transition between recombination and
fragmentation observed in the baryon/meson ratios. The
different behavior of mesons and baryons is characteristic
of the quark recombination mechanism and reproduced in all
model calculations \cite{Gr03,No03} (see Fig.~\ref{fig9}).

\begin{figure}   
\includegraphics[width=\linewidth]{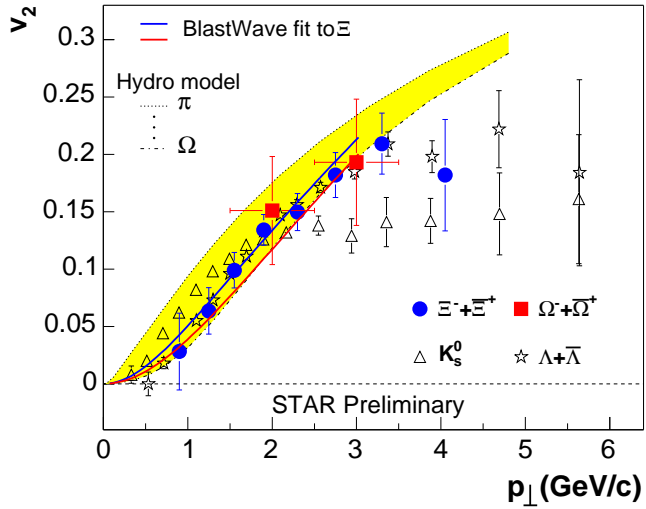}
\includegraphics[width=\linewidth]{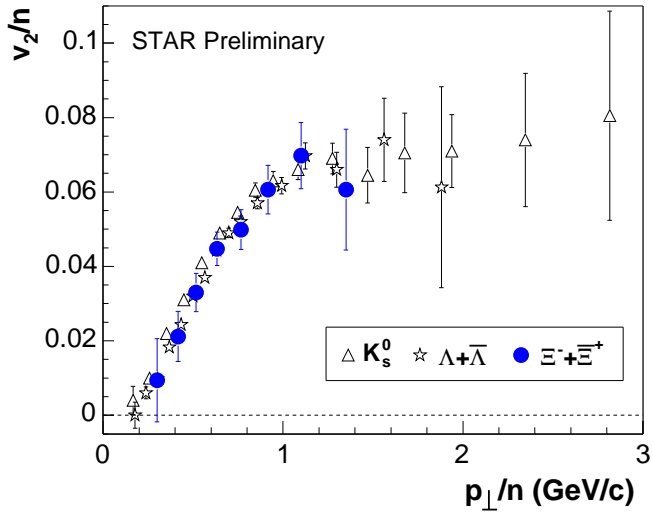}
\caption{(a) Upper panel: Elliptic flow parameter $v_2$ 
measured for various strange hadrons as function of 
transverse momentum $p_T$ (STAR data). (b) Lower panel: 
Same data, but plotted in the scaled variables $v_2/n$
and $p_T/n$, where $n=2,3$ is the number of constituent
quarks in a hadron.}
\label{fig8}
\end{figure}

\section{Summary and Outlook}

Experiments with relativistic heavy ions are allowing us, 
for the first time, to verify that a vacuum state can be 
modified under extreme conditions, causing dramatic changes 
in the properties of fundamental particles (quarks) and the
strong forces among them. Hadrons emerging by recombination
of the constituents of the hot QCD medium provide evidence
for its flavor chemical equilibration and the deconfinement
of (constituent) quarks. While flavor equilibration was
already observed in experiments at the CERN-SPS, the 
characteristic momentum dependence of the baryon/meson ratios
and the anisotropic collective flow patterns of hadrons 
have been observed at RHIC for the first time. Both features
are model independent predictions of the quark recombination
mechanism in a kinematic range where general arguments about 
the dominance of recombination apply if, and only if, a
thermally and chemically equilibrated quark plasma hadronizes.

\begin{figure}   
\includegraphics[width=\linewidth]{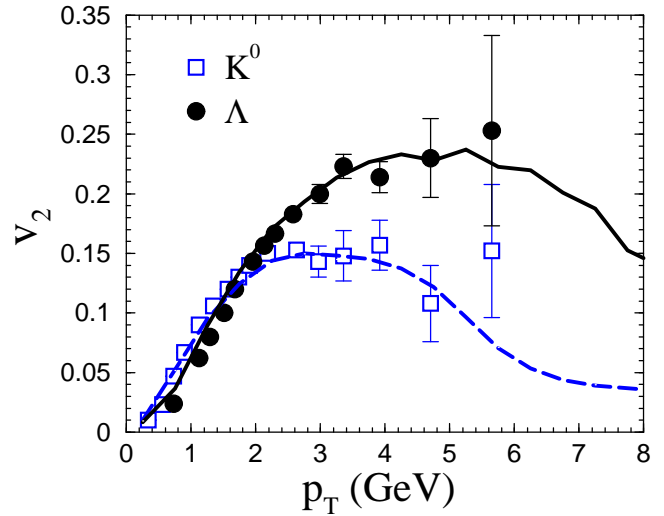}
\includegraphics[width=\linewidth]{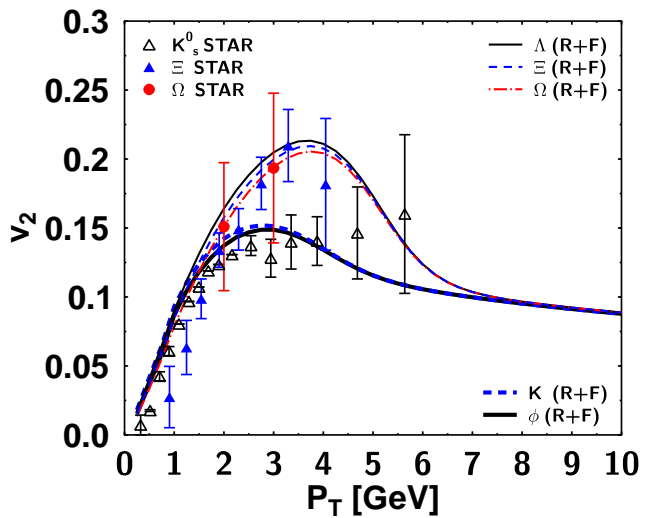}
\caption{Elliptic flow parameter $v_2$ as function of 
transverse momentum $p_T$ together with the STAR data.
(a) Upper panel: Kaons and $\Lambda$ hyperons (Greco
et al. \protect\cite{Gr03}). 
(b) Lower panel: Kaons, $\Xi$ and $\Omega$ hyperons 
(Nonaka et al. \protect\cite{No03}).}
\label{fig9}
\end{figure}

The present data and their theoretical descriptions leave 
a number of important questions unaddressed. One concerns
two-body correlations among energetic hadrons. Recombination
from a purely thermal medium only allows for quantum (HBT)
correlations among hadrons, but there is clear evidence in
the data for the persistence of jet-like correlations into
the kinematic regime where the recombination mechanism is
thought to dominate. One explanation could be that these
correlations are vestiges of the recombination of thermal
and shower partons. Another explanation could be that the
assumption of a completely thermal medium is a simplification,
and that two-parton correlations become increasingly important
with increasing $p_T$. These two explanations are likely just
two faces of the same mechanism, because correlations among
partons are probably caused by interactions of an energetic
parton with constituents of the medium.

Another important puzzle is the apparent absence of gluonic
degrees of freedom in the recombination mechanism. Does this
indicate that gluons disappear as independent degrees of
freedom as the plasma approaches the critical temperature
from above? The precipitous drop in the Debye screening mass
of a thermal plasma found in (quenched) lattice-QCD simulations 
near, but above $T_c$ (see Fig.~8 of \cite{Ka00}) would suggest 
that this is, indeed, the case. One way to explain the absence
of gluons is to argue that they fragment into quark-antiquark
pairs prior to recombination \cite{HY02}.

The presence of strong dynamical correlations (clustering) 
among the constituents of the medium prior to hadronization, 
as suggested by theoretical considerations \cite{SZ04}, would 
certainly serve to enhance the probability of recombination. 
This would be similar as the preformation of four-nucleon 
clusters facilitating the $\alpha$-particle decay of heavy 
nuclei \cite{alpha}.

\section*{Acknowledgments}

This work was supported in part by grant No. DE-FG02-96ER40945
from the U.~S.~Department of Energy. I thank Steffen Bass,
Rainer Fries, Rudolph Hwa, Che-Ming Ko, and Thorsten Renk for 
valuable comments on the manuscript and Olga Barannikova, 
Rainer Fries, Vincenzo Greco, Chiho Nonaka, Richard Seto and 
Paul Sorensen for providing figures.


\begin{thebibliography}{99}
\bibitem{others}
Evidence for the formation of a quark-gluon plasma in 
Au+Au collisions at RHIC from various other signature 
observables is discussed in various other articles in 
this volume.
\end{thebibliography}
\end{document}